\documentclass[pre,twocolumn]{revtex4}

\usepackage{epsfig}
\usepackage{amssymb,amsmath,stmaryrd,tabularx}
\usepackage{float}
\usepackage{wrapfig}
\usepackage{subfigure}
\usepackage{bm}
\usepackage{graphicx}
\newcommand{\B}[1]{{\bm{#1}}}
\begin{document}
\title{The dynamics of cracks in torn thin sheets}
\author{Yossi Cohen and Itamar Procaccia}
\begin{abstract}
Motivated by recent experiments, we present a study of the dynamics of cracks in thin sheets. While the equations of
elasticity for thin plates are well known, there remains the question of path selection for a propagating crack.
We invoke a generalization of the principle of local symmetry to provide a criterion for path selection and demonstrate qualitative agreement with the experimental findings. The nature of the singularity at the crack tip is studied with
and without the interference of nonlinear terms.
\end{abstract}
\affiliation{Department of Chemical Physics, The Weizmann Institute of Science, Rehovot 76100, Israel}
\date{\today}
\maketitle
\section{Introduction}
\begin{figure}[h]
  \includegraphics[scale=0.45]{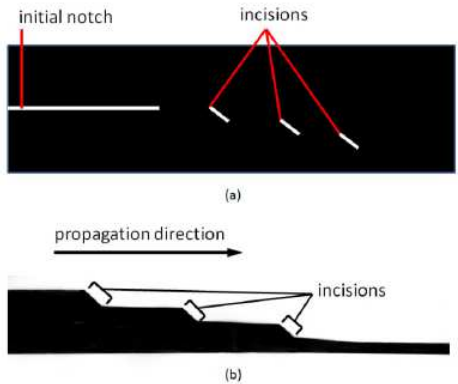}
  \caption{A typical experiment discussed in Ref. \cite{10BAB}: (a)Schematic of the sheet before tearing : a centered cut was initiated in the film with inclined incisions at $45^\circ$. (b) Scan of the film after tearing : the path appears to be stable in an asymmetric configuration.}
  \label{incisions}
\end{figure}

\begin{figure}[h]
  \includegraphics[scale=0.45]{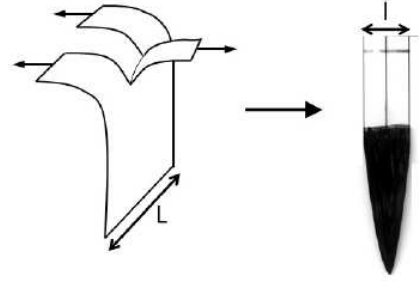}
  \caption{Another typical experiment discussed in Ref. \cite{10BAB}: Two cracks propagating simultaneously. The middle part of the film forms a tongue-like shape and detaches itself.}
  \label{tounge}
\end{figure}

This paper is motivated by some recent experiments in which thin sheets were torn under an out-of-plane shear mode
(known also as mode III) \cite{10BAB}. The experiments tested a few initial crack configurations, including an initial notch
placed symmetrically in the middle of the thin sheet, but also richer configurations as shown in Fig. \ref{incisions}
and Fig. \ref{tounge}. The latter configuration is interesting in having two interacting cracks which appear to attract each other until a `tongue' is separated from the mother sheet when the cracks coalesce. These experiments pose a challenge for theory since it is not a-priori known what is the criterion for path selection for cracks in thin plates, or, in other words, where should the crack turn under the action of a given stress field. The aim of this paper is to address this question and to show that a proper combination of elasticity theory in thin plates and a reasonable extension of the principle of local symmetry \cite{74GS} results in crack propagation that is in a reasonable agreement with the experimental results.

The structure of this paper is as follows: in Sect. \ref{laplace} we remind the reader the well known theory of regular mode III fracture in a thick plate, and the modifications required for thin plates.
In Sect. \ref{singular} we discuss the nature of singular stress field at the tip of crack in a thin plate. Sect \ref{nonlinear} considers the modifications introduced by the large deflections that are typical to the experiment
under study. In Sect. \ref{numerics} we describe our numerical procedures and compare the singular solution
of the stress field to the theoretical expectations, including the influence of nonlinear terms. Next we consider
the criterion of path selection and the resulting crack trajectories for various initial conditions.
Comparison with the experiment are offered. Lastly, in Sect. \ref{conclusions} we summarize the paper and offer
conclusions.

\section{Mode III in plates}
\label{laplace}
\subsection{Thick Plates}
Imagine a crack which is situated along a symmetry plane (in the middle of a thick plate) subject to anti-symmetric distributed loads, directed in parallel with the crack edge. For a thick plate, we assume that the elastic displacements in the body are
\begin{eqnarray}
u&=&0, v=0, \nonumber  \\
w&=&w(x,y).
\end{eqnarray}
The displacements in the $xy$ plane are denoted by $u,v$, respectively, and the displacement in the $z$ direction is denoted as $w$. Here and throughout this work we use the Lagrangian convention in which displacements, strains and stresses are computed in the equilibrium position of a material point. According to Hooke's law, the components of the stress tensor are \cite{61BC}
\begin{eqnarray}
\sigma_{xx}=\sigma_{yy}=\sigma_{zz}=\sigma_{xy}=0  \nonumber \\
\sigma_{yz}=\mu\frac{\partial w}{\partial y}, \sigma_{xz}=\mu\frac{\partial w}{\partial x}.
\end{eqnarray}
where $\mu=E/2(1+\nu)$ is the shear modulus, $E$ Young's modulus and $\nu$ Poisson's ratio.

In a cylindrical coordinate system $r,\varphi,z$, where $r=0$ coincide with the crack edge and $\varphi=\pm\pi$ with the crack faces, the stresses become
\begin{equation}
\sigma_{rz}=\mu\frac{\partial w}{\partial r}, \sigma_{\varphi z}=\mu\frac{1}{r}\frac{\partial w}{\partial \varphi}.
\label{str}
\end{equation}
The condition of equilibrium is
\begin{equation}
\frac{\partial \sigma_{rz}r}{\partial r} + \frac{\partial \sigma_{r\varphi}}{\partial \varphi} = 0
\end{equation}
This equation can be written in a compact and coordinate-invariant form of the Laplace equation
\begin{equation}
\triangle w=0.
\end{equation}
The Laplace equation is solved by any harmonic functions, with a specific solution selected by the boundary conditions. The crack faces are assumed to be traction free,
\begin{equation}
\frac{\partial w}{\partial \varphi}=0 \ for \ \varphi=\pm \pi.
\label{boun}
\end{equation}
Due to the predominant anti-symmetry the harmonic solution that satisfied the boundary condition (\ref{boun}) can be expressed as the Fourier series \cite{99B},
\begin{equation}
w(r,\varphi)=\sum_{n=1,3,..}(A_nr^{n/2} + B_nr^{-n/2})\sin\left(\frac{n\varphi}{2}\right)
\label{fouw}
\end{equation}
where $A_n$ and $B_n$ are constants. Applying this to Eq. (\ref{str}) yields the general form of the stresses
\begin{eqnarray}
\sigma_{rz}=\sum_{n=1,3,..}(a_nr^{n/2-1} + b_nr^{-n/2-1})\sin\left(\frac{n\varphi}{2}\right) \nonumber \\
\sigma_{\varphi z}=\sum_{n=1,3,..}(a_nr^{n/2-1} + b_nr^{-n/2-1})\cos\left(\frac{n\varphi}{2}\right)
\end{eqnarray}
Barenblatt and Cherepanov \cite{61BC} showed that the stress $\sigma_{\varphi z}$ for small $r$  has a singularity of the type $r^{-1/2}$, and close to the crack tip it can be represented as
\begin{equation}
\sigma_{\varphi z}=\frac{K_{III}}{\sqrt{2\pi r}}\cos\left(\frac{\varphi}{2}\right)
\label{taulap}
\end{equation}
where $K_{III}$ is the \emph{stress intensity factor} which expresses the amplitude of the singularity.

\subsection{The Kirchhoff Thin Plate Theory}
 \label{Kirchhoff}
 In thin plates the result Eq. \ref{taulap} is no longer true.
 In order to analyze the correct stress singularity near the crack edge, we need to obtain the equation for the deflection $w$ (displacement in the $z$-axis) for a thin plate \cite{86LL,T40}. We consider a plate of uniform thickness, equal to $h$, and take the $xy$ plane at the middle plane of the plate where $z=0$.  For a small deflection the displacements $u,v$ vanish in this plane. However, the bending causes a small deformation along the $z$-axis, so the displacements become
\begin{equation}
u= -z \frac{\partial w}{\partial x} \ , \quad
v= -z \frac{\partial w}{\partial y} \ .
\end{equation}

These create bending moments and a twisting moment that act on the plate elements and can be expressed as \cite{T40}
\begin{eqnarray}
M_x&=&\int_{-\frac{h}{2}}^{\frac{h}{2}}\sigma_{xx} zdz=-D\left(\frac{\partial^2 w}{\partial x^2} + \nu\frac{\partial^2 w}{\partial y^2}\right) \nonumber \\
M_y&=& \int_{-\frac{h}{2}}^{\frac{h}{2}}\sigma_{yy}zdz=-D\left(\frac{\partial^2 w}{\partial y^2} + \nu\frac{\partial^2 w}{\partial x^2}\right) \nonumber \\
M_{xy}&=& \int_{-\frac{h}{2}}^{\frac{h}{2}}\sigma_{xy}zdz=D(1-\nu)\frac{\partial^2 w}{\partial x\partial y}
\end{eqnarray}
where $D=Eh^3/12(1-\nu^2)$ is called \emph{the flexural rigidity} of a plate.

Substituting these expressions in the equation of equilibrium, the equation for the deflection $w$ for pure bending becomes
\begin{equation}
\triangle^2 w=0
\label{bh}
\end{equation}

In addition to the bending moments $M_x$ and $M_y$ and the twisting moment $M_{xy}$, there are vertical shearing forces acting on the sides of the element.
\begin{equation}
Q_x=\int_{-\frac{h}{2}}^{\frac{h}{2}}\sigma_{xz}dz,\ \ \ \ Q_y=\int_{-\frac{h}{2}}^{\frac{h}{2}}\sigma_{yz}dz.
\end{equation}

These forces can be expressed in cylindrical coordinates as \cite{T40}
\begin{eqnarray}
Q_r&=&-D\frac{\partial}{\partial r}(\triangle w), \label{qr} \\
Q_t&=&-D\frac{\partial(\triangle w)}{r \partial\varphi}.
\label{qt}
\end{eqnarray}

For a crack in mode III, the crack propagation is controlled by the transverse shear stress  $\sigma_{\varphi z}$ \cite{61BC,09HK}. Since it varies along the $z$ axis, it is more common to take the sum of it along the thickness of the plate Eq.\eqref{qt}. In the next section we obtain the shear force $Q_t$, and calculate the dominant terms close to the crack edge.

\section{Singularity of the Shear Force in Thin Plates.}
\label{singular}
Imagine a crack cutting a thin plate along a symmetry plane, subject to anti-symmetric load. The crack tip is at $r=0$ and the crack faces are at $\varphi=\pm\pi$.
For small bending, the deflection $w$ is a solution of the bi-harmonic equation Eq.\eqref{bh}. Assuming anti-symmetry, $w$ can be expressed as the Fourier series:
\begin{equation}
w=\sum_n b_n(r)\sin(n\varphi)
\label{FTw}
\end{equation}
where $b_n(r)$ is defined by the boundary condition and the initial load.

The crack faces are assume to be entirely free. Thus, along this edge there are no bending nor twisting moments and also there is no vertical shearing forces \cite{T40}, i.e.
\begin{equation}
M_\varphi=M_{r\varphi}=Q_t=0 \ for \ \varphi=\pm\pi
\label{freebc}
\end{equation}
Kirchhoff \cite{50K,T40} proved that two boundary conditions are sufficient for the determination of the deflection $w$. Thus two on the above conditions are added together; in cylindrical coordinates the transformed conditions at $\varphi=\pm\pi$ read
\begin{eqnarray}
M_\varphi &=&-D\left(\frac{1}{r}\frac{\partial w}{\partial r} + \frac{1}{r^2}\frac{\partial^2w}{\partial\varphi^2}+\nu\frac{\partial^2 w}{\partial r^2}\right)=0
\label{BCm}\\
V_\varphi &=&\left(Q_\varphi - \frac{\partial M_{r\varphi}}{\partial r}\right) \label{BCv} \\
&=&\!\!-D\left[\frac{1}{r}\frac{\partial\triangle w}{\partial\varphi}\! + (1-\nu)\frac{1}{r}\left(\frac{\partial^3 w}{\partial r^2\partial\varphi} -\frac{1}{r}\frac{\partial^2 w}{\partial r\partial\varphi}\right)\right]=0 \nonumber.
\end{eqnarray}
The first boundary condition is satisfied identically when the Fourier series of Eq. (\ref{FTw}) is substituted in Eq. \eqref{BCm} at  $\varphi=\pm\pi$. Repeating this with the second boundary condition \eqref{BCv} results in
\begin{equation}
-D\frac{1}{r}\left[(2\!-\!\nu)\frac{\partial^2b_n}{\partial r^2}\!-\!\nu\frac{1}{r}\frac{\partial b_n}{\partial r}\! -\! \frac{n^2}{r^2}b_n\right]n\cos(\pm\pi n)=0
\end{equation}
This equation becomes zero at $\varphi=\pm \pi$ when $n=1/2+m$ and $m\epsilon\mathbb{Z}$. Thus, $w$ can be expressed as
\begin{equation}
w=\sum_{n=1,3,..} b_n(r)\sin(\frac{n}{2}\varphi)
\end{equation}
substituting this in Eq.\eqref{bh}, and the general solution of the deflection $w$ takes the form of the following series:
\begin{equation}
w\!=\!\!\sum_{n=1,3,..}\!\! \left(\!A_nr^{\frac{n}{2}}\!+\!B_nr^{-\frac{n}{2}}\!+\!C_nr^{2-\frac{n}{2}}\!+\!D_nr^{2+\frac{n}{2}}\right)\!\sin(\frac{n}{2}\varphi)
\label{wf}
\end{equation}
where $A_n,B_n,C_n$ and $D_n$ are constants.

The stress near the crack tip will be defined by the shear force $Q_t$ which relates to the anti-plane shear stress $\tau_{\varphi z}$. From equations \eqref{qt} and \eqref{wf}, it follows that the shear force can generally be represented by an expansion of the form
\begin{equation}
\begin{aligned}
Q_t=\!\!\sum_{n=1,3,..}&\!\!-D\left[C_n(4-2n)r^{-\frac{n}{2}-1}+D_n(4+2n)r^{\frac{n}{2}-1}\right] \\
&\times\frac{n}{2}\cos(\frac{n}{2}\varphi)
\end{aligned}
\end{equation}
notice that the first two terms of Eq.\eqref{wf} give zero contribution to the shear force, since both of them are harmonic solutions for the Laplace equation.
Now, taking the leading terms for $n=1$, and we get that close to the crack tip the two dominant terms are
\begin{equation}
Q_t=-D\left(c_1r^{-\frac{3}{2}} + d_1r^{-\frac{1}{2}}\right)\cos(\frac{\varphi}{2}).
\end{equation}

This result shows, that while in the standard mode III the inverse square-root is the dominant term close to the crack tip, for thin plates the dominant term scales as $r^{-\frac{3}{2}}$ \cite{98ZHCH}. We stress that this highly singular
term is the result of linear theory. The numerical results presented below show that upon increasing the bending of the plate, this linear singularity is weakened due to non-linear effects.

\section{Large deflection of thin plates.}
\label{nonlinear}
The theory of thin plates is applicable only for small deflections. In most cases it is applicable when the deflection $w$ is small compared to the thickness of the plate \cite{86LL}. Thus, to relate to the experiments of Ref. \cite{10BAB} we must consider also large deflection effects. While for small deflections the basic assumption was that the middle surface remains undeformed, here its deformation can not be neglected. Due to the large deflection, the quadratic terms in the strain tensor can not be omitted. For clarity of presentation we use now the notation $\B u$ for the in-plane displacement with $u_\alpha$ and $u_\beta$ replacing $u$ and $v$ of the previous sections; we leave $w$ to denote the $z$-component of the displacement. The strain tensor becomes
\begin{equation}
\varepsilon_{\alpha\beta} = \frac{1}{2}\left(\frac{\partial u_\alpha}{\partial x_\beta} + \frac{\partial u_\beta}{\partial x_\alpha}\right)+\frac{1}{2}\frac{\partial w}{\partial x_\alpha}\frac{\partial w}{\partial x_\beta}.
\end{equation}
The quadratic terms of $u_\alpha$ is neglected by assuming small deformation, but the same cannot be done with the derivative of $w$.
Taking the nonlinear term into account, the final equation for the deflection becomes \cite{86LL,T40}
\begin{equation}
D\triangle^2 w - h\frac{\partial}{\partial x_\beta}\left(\sigma_{\alpha\beta}\frac{\partial w}{\partial x_\alpha}\right)=0 .
\label{nlw}
\end{equation}
where $\sigma_{\alpha\beta}$ is the in plane stress acting in the middle plane of the plate. Assuming that there is no body forces or tangential forces acting in those direction we obtain the following equations of equilibrium:
\begin{equation}
\frac{\partial \sigma_{\alpha\beta}}{\partial x_\beta}=0.
\label{nls}
\end{equation}

The three equations \eqref{nlw} and \eqref{nls} (Eq. \eqref{nls} is actually two equations), can be reduced to two equations by introducing the Airy stress function $\chi$ defined by
\begin{equation}
\sigma_{xx}=\frac{\partial^2\chi}{\partial y^2}, \  \sigma_{yy}=\frac{\partial^2\chi}{\partial x^2}, \ \tau_{xy}=-\frac{\partial^2\chi}{\partial x\partial y}.
\label{chi}
\end{equation}
Substituting in Eqs. \eqref{nlw} and \eqref{nls} we obtain the complete system of equations for large deflection of a thin plate
\begin{eqnarray}
&D\triangle^2 w\! - \!h \left(\frac{\partial^2\chi}{\partial y^2}\frac{\partial^2 w}{\partial x^2}\! + \! \frac{\partial^2\chi}{\partial x^2}\frac{\partial^2 w}{\partial y^2}\! -\!2 \frac{\partial^2\chi}{\partial x\partial y}\frac{\partial^2 w}{\partial x\partial y}\right)\!=\!0  \label{fvk1}\\
&\triangle^2\chi+E\left\{\frac{\partial^2 w}{\partial x^2}\frac{\partial^2 w}{\partial y^2}-\left(\frac{\partial^2 w}{\partial x\partial y}\right)^2\right\}=0 \label{fvk2}
\end{eqnarray}
These equations are known as \emph{F\"{o}ppl von-K\'{a}rm\'{a}n equations}. These equation are non linear and can not be solved exactly even in very simple cases.

It should be noticed that the bending term (the first term of Eq.\eqref{fvk1}) has a higher power of $h$ than the stretching term ($D\sim h^3$). Thus, for a thin plate subject to large deformation, the bending term can be neglected in comparison to the stretching term and Eq.\eqref{nlw} becomes
\begin{equation}
h\sigma_{\alpha\beta}\frac{\partial^2w}{\partial x_\alpha\partial x_\beta}=0
\end{equation}
In case of isotropic stretching, the stress components become the same in all directions, the equation for the deflection becomes \cite{86LL}
\begin{equation}
\triangle w=0
\end{equation}
which is similar to the case of mode III for thick plates.
The upshot of the last remarks is that for small deflection of a thin plate in mode III we  expect that the anti-plane shear stress will result in  a tip-singularity of the form $r^{-\frac{3}{2}}$. However, upon increasing the deflection, and therefore increasing the in-plane stresses near the crack edge, this highly singular term should weaken, turning eventually to a standard $r^{-\frac{1}{2}}$ singularity. We will see that our numerical results support this expectation.
\section{Numerical Results}
\label{numerics}
\subsection{The crack tip singularity.}
\begin{figure}
\hskip -1.5 cm
  \includegraphics[scale=0.37]{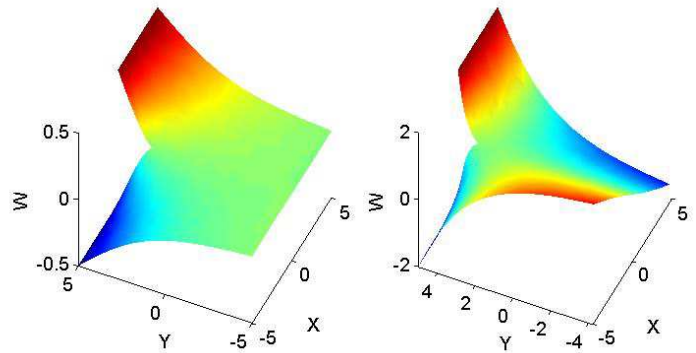}
  \caption{Color online. Left panel: the bi-Laplace solution for small deflection on thin plates under mode III conditions. Right side: The solution of the FvK equation which include the non-linear terms. The colors indicate the size of the deflection $w$.}
  \label{ps}
\end{figure}

\begin{figure}
  \includegraphics[scale=0.6]{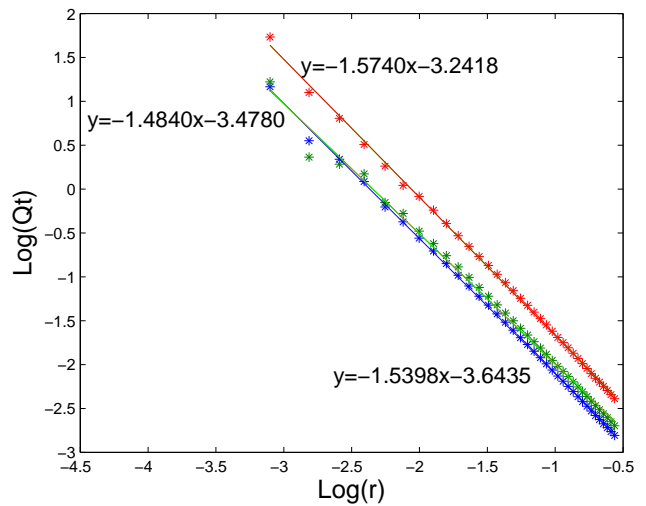}
  \caption{Color online. A log-log plot of the shear force as a function of the distance from the crack. The three sets of data are for three different crack lengths. The best linear fits indicate the theoretical expectation that the stress close to the crack tip scales like $r^{-3/2}$. }\label{ds}
\end{figure}

\begin{figure}[h]
  \includegraphics[scale=0.6]{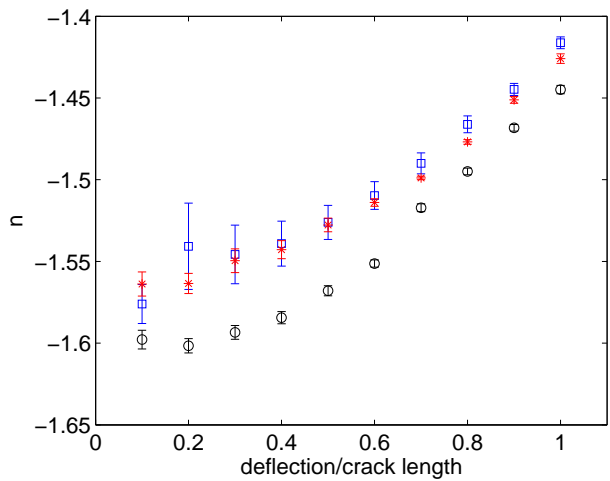}
  \caption{The index of the shear force singularity $r^n$ as a function of the amount of bending at mode III for several crack length. Increasing the bending creates an in-plane stress and weakens the singularity of the shear force at the crack tip.}
  \label{np}
\end{figure}

In order to analyze the stress field near the crack tip, we consider a system consisting of a rectangle thin elastic plate [$L\times L \times h$; $h/L\sim0.05$]. An initial cut of length $l$ was created at the middle of the upper edge. One side of the upper edge of the plate $\{-L/2<x<0; y=L/2\}$ is bent to one direction and the other side $\{0<x<L/2; y=L/2\}$ to the opposite direction, cf. Fig. \ref{ps}. The other edges, including the crack's faces, are considered as free (cf. Eqs. \ref{freebc}). We developed a solver using the Galerkin finite elements method on a triangular grid to calculate numerically the stress field and the shape of the plate. For the solution of the non-linear equations \eqref{fvk1},\eqref{fvk2}, we resort to a combined incremental-iterative (Newton-Raphson) procedure.
The solution of the shear force as a function of the distance from the tip for the bi-Laplace equation \eqref{bh} for a small deflection is shown in Fig. \ref{ds}. We obtain that the singularity exponent of the shear stress close to the tip for several initial crack length is close to the theoretically expected value of -3/2.
Next we investigate what happens upon increasing the bending of the thin plate. Close to the crack tip, the Gaussian curvature (the second term of Eq.\eqref{fvk2}) is non zero and therefore an in-plane stress begins to accumulate. The bi-laplace Eq. \eqref{bh} does not represent the situation any more;  We must turn to the FvK Eqs. \eqref{fvk1},\eqref{fvk2}.
The difference between the solutions of these equations can be clear to the eye, as we demonstrate in Fig. \ref{ps}.
In the left panel the solution of the Kirchhoff plate theory does not involve the creation of additional in-plane
stresses. In the right panel the nonlinear theory creates in-plane stresses at the crack tip, resulting in a sizeable
distortion of the plate shape ahead of the crack. This visual effect is also seen in the weakening of the stress singularity at the crack tip. In Fig. \ref{np} we show, for several crack length, the singularity of the shear force. Note that for large bending the singularity of the shear force is weakened.
\subsection{Crack Propagation}
The solution of the elastic problem, linear or nonlinear, does not contain a recipe how to advance the crack in a given stress field. For a precisely anti-symmetrical mode of tearing, the crack has no reason not to develop in a precisely straight trajectory. However, breaking the symmetry causes the crack to propagate in a different direction, and this direction is not known a-priori even when we know the stress field. We propose that in the present situation one can follow the strategy of Ref. \cite{61BC} which for the Laplacian problem introduced a subdominant term to Eq.\eqref{taulap}. This term breaks the symmetry
\begin{equation}
\tau_{\varphi z}=\frac{K_{III}}{\sqrt{2\pi r}}\cos\left(\frac{\varphi}{2}\right) - \mu A_2\sin(\varphi)
\label{taubs}
\end{equation}
The hypothesis of Ref. \cite{61BC} was that a crack under longitudinal shear propagates along a direction that annuls $A_2$, making the stress distribution close to the crack tip symmetrical about the crack direction. For thin plates we can also examine the subdominant term that breaks the symmetry for the deflection $w$. This term is proportional to $\cos({\varphi})$,
\begin{equation}
w_n=\sum\left(\tilde{a}_nr^n + \tilde{b}_nr^{-n} + \tilde{c}_nr^{2-n} + \tilde{d}_nr^{2+n}\right)\cos(n\varphi).
\end{equation}
substitute this in Eq.\eqref{qt}, and the first two terms vanish, being solutions of the Laplace equation. For $n=1$ the third term also vanishes, and the first subdominant term that breaks the symmetry for thin plates comes from the last, $r^3$ term. Substituted in Eq.\eqref{qt}, it becomes identical to the term found in the Laplacian case Eq.\eqref{taubs}
\begin{equation}
Q_t= -D\left(c_1r^{-\frac{3}{2}} + d_1r^{-\frac{1}{2}}\right)\cos(\frac{\varphi}{2}) + \mu A_2\sin(\varphi).
\label{Qt}
\end{equation}
We thus propose that the `principle of local symmetry' in the present case dictates propagating the crack such as to
annul $A_2$.
The generally accepted principle of local symmetry for crack propagation in mode I and II \cite{74GS} thus has a general analog for mode III as the condition $A_2=0$. Of course, whether the crack propagates at all is dictated by the analog of the Griffith criterion, i.e. the energy release rate into the crack tip region from the elastic field must be larger than the dissipation involved in the creation of free surfaces. Note that in the standard case of 1/2 singularity one
possesses analytic estimates of the energy release rate and by assuming a constant energy dissipation per unit area
of free surface one can estimate the critical length of a crack. In the present case the analytic estimates of the
coefficients $c_1$ and $d_1$ in Eq. (\ref{Qt}) are not available; their calculation requires further work, solving
the bi-Laplace equation analytically. This calculation is beyond the scope of the present paper, but is planned
for a future publication
\subsection{Single crack trajectories with the principle of local symmetry}
\begin{figure}
\subfigure[Laplacian field]
{\includegraphics[scale=0.28]{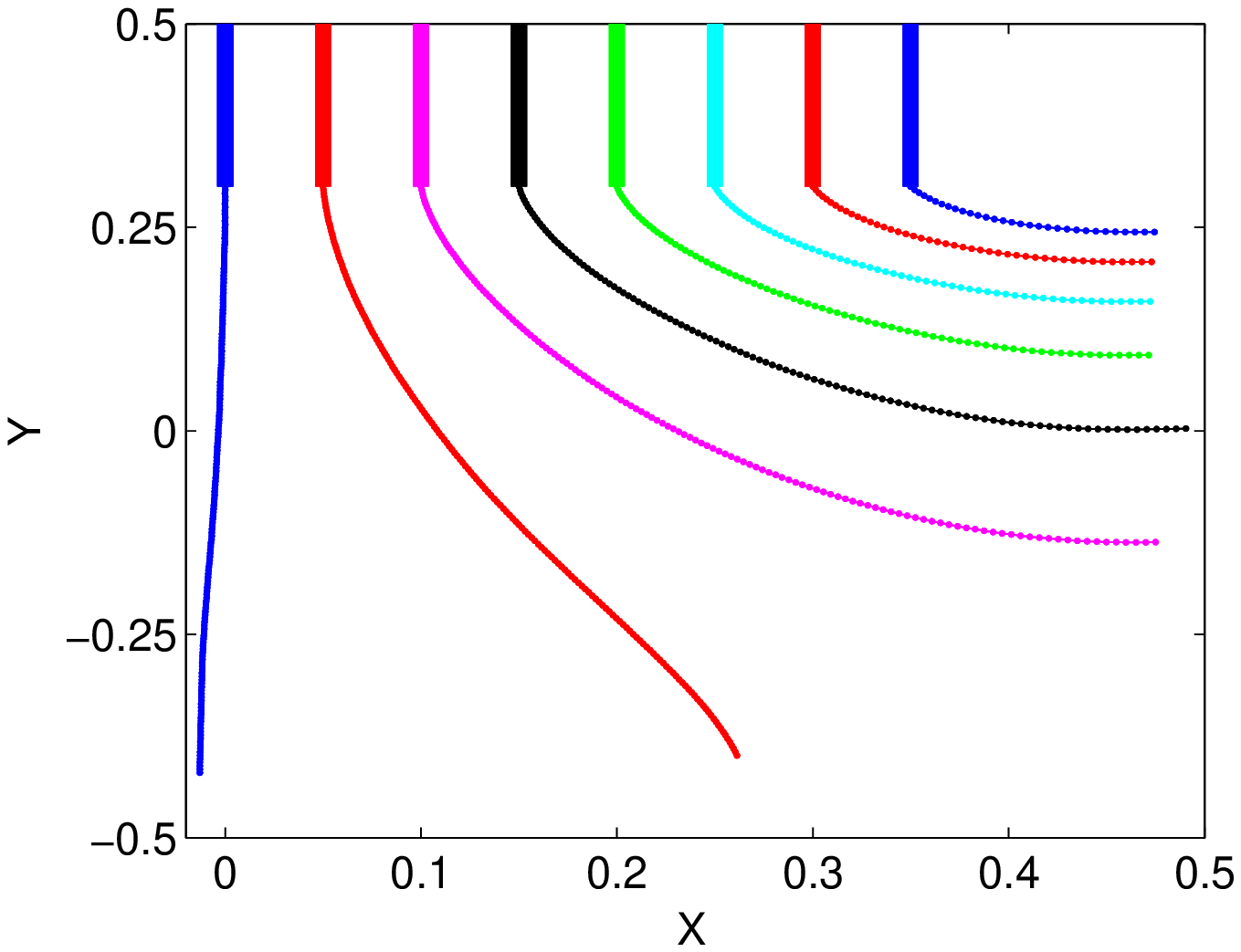}\label{cp1}}
  \subfigure[FvK solution]{
  \includegraphics[scale=0.28]{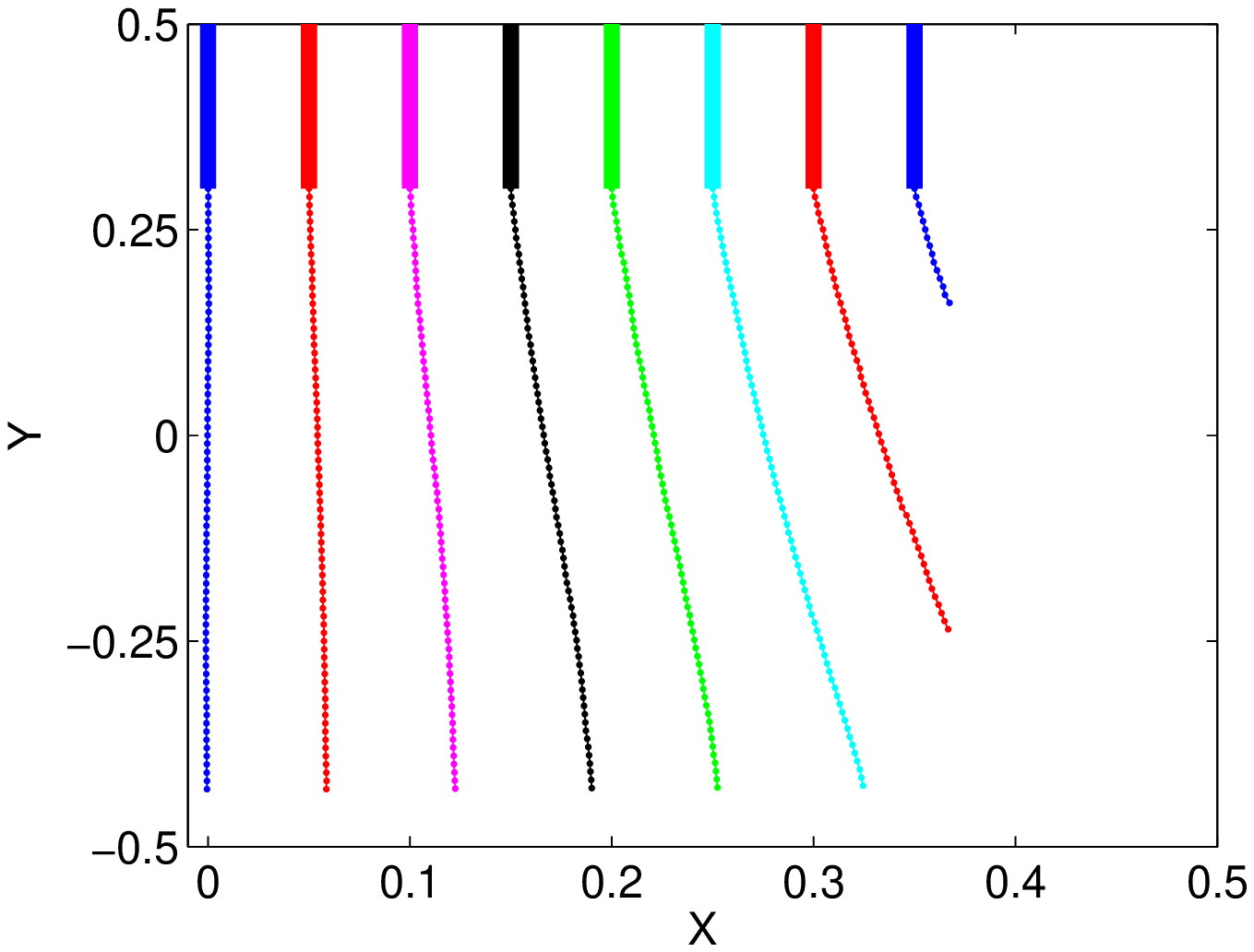}\label{cp2}
  }
  \caption{Color online. Crack paths for the \emph{normal} mode III (upper panel) and for the thin plate in mode III (lower panel). The thick line shows the initial crack. Note when the crack started in the middle of the upper edge, $x=0$, it continues straight for both cases, up to numerical inaccuracies.}\label{cp}
\end{figure}

We consider a plate of size $1\times1 \ \{-0.5<x<0.5, -0.5<y<0.5\}$, with a small cut created in the upper edge ($y=0.5$) at different locations parallel to the boundary. One side of the upper edge is bent to one direction and the other to the opposite direction (Fig \ref{ps}). We chose the Young modulus $E=50$ and the Poisson ratio $\nu=0.3$. We calculate the stress field for two cases: one for the Laplacian field, the \emph{normal} mode III, and the second, for a thin plate, Eqs (\ref{fvk1}),(\ref{fvk2}) .To decide whether the crack should evolve or not we must choose a critical value of $c_1$ in Eq. (\ref{Qt}) above which the crack must develop. To do so we computed the value of $c_1$ where the deflection equals
the length of the initial crack, and chose this value of $c_1$ to be the critical value $c_1^*$. This procedure is
equivalent to choosing the energy dissipation per unit area of free surface when the Griffith criterion is employed.
In subsequent crack development we maintained the critical value of $c_1$ invariant. Thus the crack evolves according to criterion $c_1> c_1^*$ and the principle of local symmetry. Whenever the value of $c_1$ was below $c_1^*$, we increased the bending quasi-statically by bringing the deflection closer to the crack's tip. Not surprisingly, we observed that a crack that was initiated in the middle of the upper edge $x=0$ always continues straight, Fig. \ref{cp}. However, when the initiation breaks the symmetry by starting the crack right to the center, the crack tends to turn to the right boundary. This tendency is stronger in the Laplacian case, where the crack curves sharply toward the right boundary. For the thin plates, the crack shows more stability by creating a smaller deviation from the straight line, and thus reaching the bottom in the present geometry.

\begin{figure}
\subfigure[Inclined cracks]{
  \includegraphics[scale=0.27]{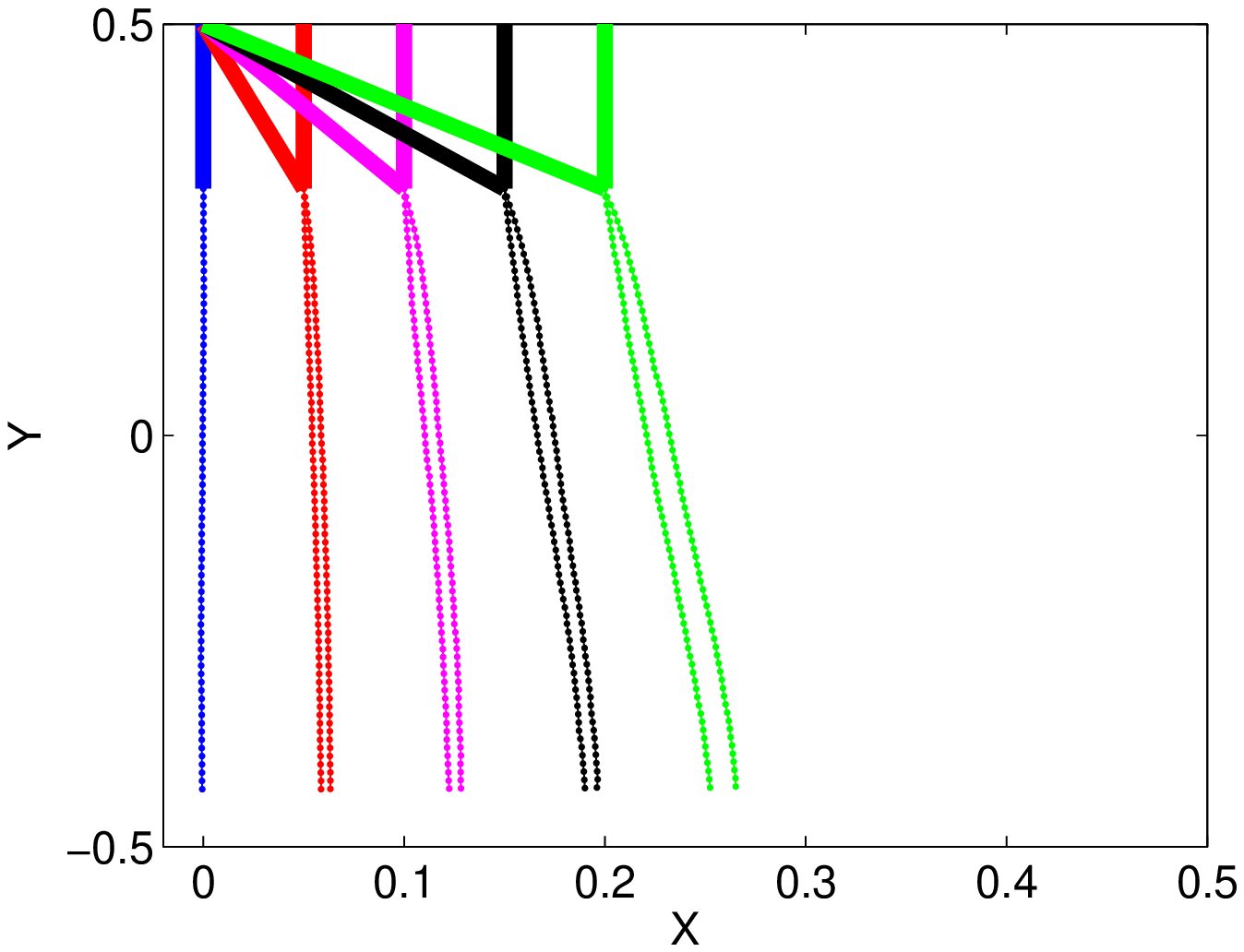}\label{ic1}
  }
  \subfigure[Thin sheet with incisions]{
  \includegraphics[scale=0.27]{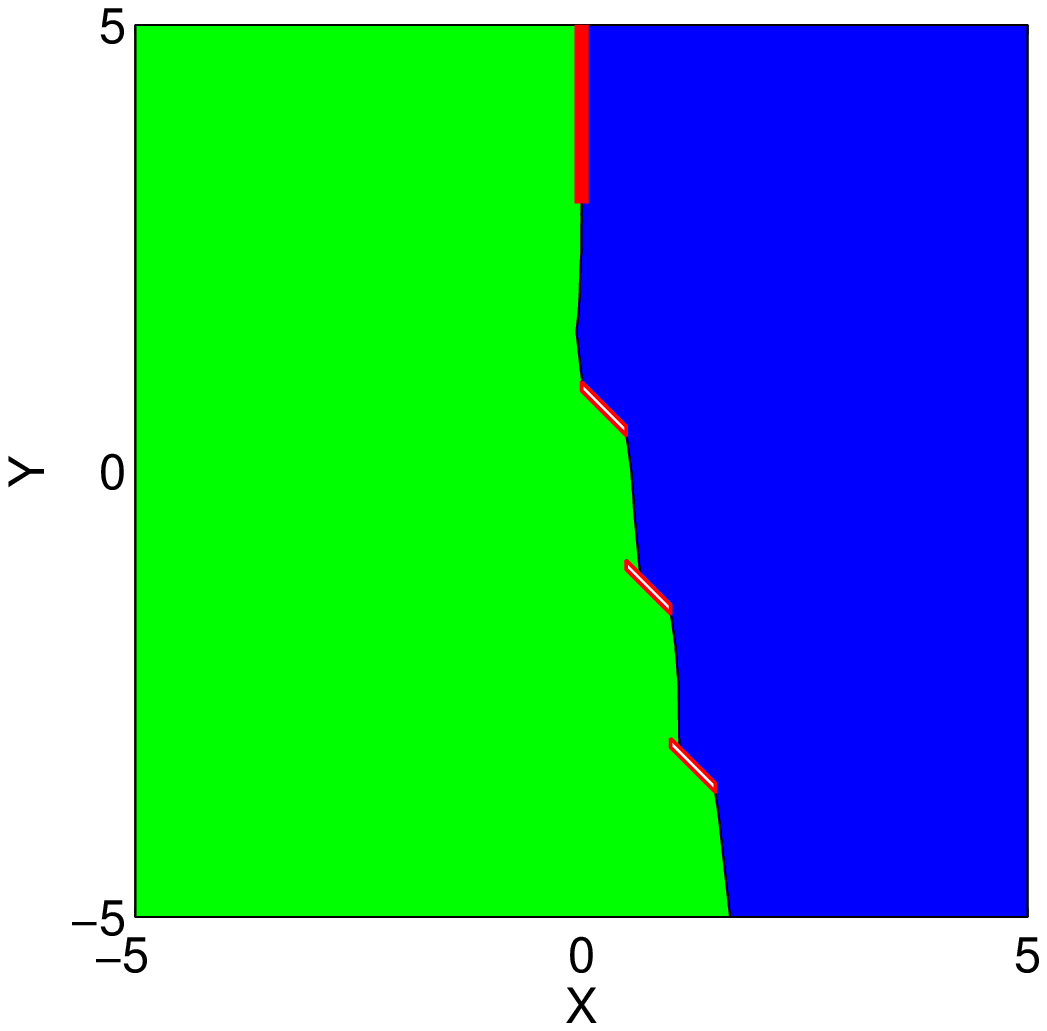}\label{ic2}
  }
  \caption{Color online. Upper panel: a comparison of the crack path of an inclined crack that starts at the center of the upper edge with an angle other than a right angle, and a crack that started with a right angle but not at the center of the edge. Lower panel: the calculated crack path with incisions of $45$ degrees inside the plate (red parallelograms).}\label{ic}
\end{figure}

Next we investigated the trajectories followed by an inclined crack. We initiate these with an angle with respect to the upper edge, such that a `straight' initial crack has a right angle. In Fig. \ref{ic1} we show that such initial cracks turn sharply towards the same trajectory that is followed by a `straight' crack whose tip ends at the same point.
This result indicates a surprising lack of memory for past history. The principle of local symmetry forces the inclined initial crack to turn sharply such as to agree almost exactly with the future trajectory of a straight initial crack whose tip is at the same point. Last, we compare the numerical solution to the experimental configuration Fig. \ref{incisions}.
We mimicked a similar set of incisions, and found that indeed the crack straightens up between the incisions as reported in the experiment.  Even when creating incisions inside the sheet at an angle of $45^o$, the crack turns and continue almost in a straight direction (Fig. \ref{ic2}).
\begin{figure}
\includegraphics[scale=0.4]{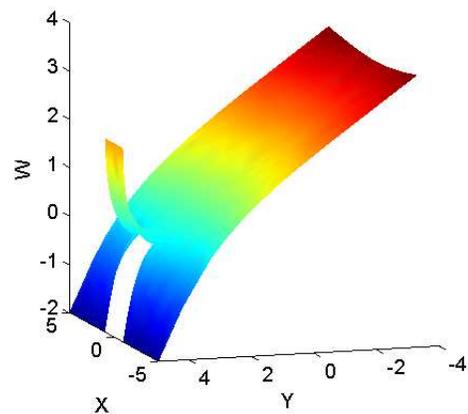}
\caption{Color online: the solution for the deflection $w$ of the FvK equations for two cracks in mode III. The color code measures the deflection $w$. Notice that towards its end the sheet curves as a result of the coupling of the bending and the in-plane stress in the FvK equation.}
\label{dc1}
\end{figure}
\begin{figure}
\includegraphics[scale=0.5]{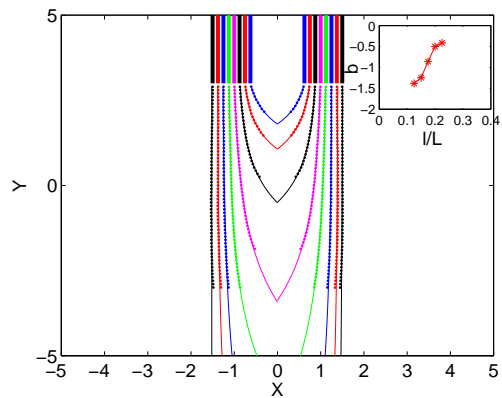}
\caption{Color online. Numerical results for two cracks trajectories. Thick straight lines are the initial cracks, the dots are the results of the simulation; the thin line represent an exponent fit to the crack shape. In the inset, the numerical factor in the exponential function of the distance between the cracks (normalized to the width of the sheet).}
\label{dc2}
\end{figure}

\subsection{Interacting Cracks}
The experiment showed in Ref. \cite{10BAB} showed that while a single crack is relatively stable, when two cracks propagate simultaneously in parallel to each other, they curve toward the center to create a tongue-like shape. In this simulation, we initiate two cracks symmetrically around the central axis, and bent the center to one direction, and the left and the right side to the other direction, Fig \ref{dc1}. The cracks evolve according to the conditions described in the last section. We allow both cracks to propagate at each step. By changing the distance between the cracks, we found that there exist a distance between the cracks at which both continue almost straight without attracting each other, cf. Fig \ref{dc2}. Theoretically one could expect that this minimal distance is 1/3 of the sample width, where the attraction to the edges cancels exactly the attraction between the cracks.  When the distance between the cracks is smaller, they attract each other. We found a fair analytic fit to the path of the cracks in the form of an exponential function $y=a(1-exp(bx))$, where $x$ is distance from the intersection point of the two cracks, $y$ is the distance between the two cracks, and $a,b$ are constant. In the inset of Fig. \ref{dc2} we present the numerical factor $b$ ; when the distance between the crack is larger from $l/L>0.225$ we can not find an intersection point and the numerical factor $b$ is close to zero.

\section{Conclusions}
\label{conclusions}
In summary, we offered a criterion for path selection for cracks in thin plates. The criterion appears to generate cracks that agree with those seen in the experiment. It therefore appears that numerical solutions can provide reliable predictive tools for the dynamics of cracks in think plates in the Mode III tearing mode. In addition we analyzed the stress field for mode III, for thick and thin plates, and showed that the singularity of the stress field close to the crack tip is stronger in the latter case, but it weakens when small in-plane deformations commence.

\acknowledgements
We are grateful to Mokhtar Adda-Bedia and Arezki Boudaoud for sharing with us their experimental results and for hosting Y.C. in Paris. We benefitted from discussions with Eran Bouchbinder. This work has been supported in part by the German-Israeli Foundation, the Israel Science Foundation and the Minerva Foundation, Munich, Germany.


\begin{thebibliography}{99}
\bibitem{10BAB}
E. Bayart and M. Adda-Bedia and A Boudaoud, ENS preprint.

\bibitem{74GS}
R. V. Goldstein and R. L. Salganik, Int. J.Fract. {\bf 10}, 507 (1974).

\bibitem{61BC} G. I. Barenblatt and G. P. Cherepanov, PMM {\bf 25}, 1110 (1961).

\bibitem{99B}
K. B. Broberg, \textit{Crack and Fracture}, (Academic Press, London) 1999.

\bibitem{86LL}
L. D. Landau and E. M. Lifshitz, \textit{Theory of Elasticity}, (Pergamon Press, Oxford) 1986.

\bibitem{T40}
S. Timoshenko, \textit{Theory of plates and Shells}, (McGraw-Hill Book Company, New York) 1940.

\bibitem{09HK}
V. Hakim and A. Karma, J Mech Phys Solids {\bf 57}, 342 (2009).


\bibitem{50K}
G. Kirchhoff, J. Crelle {\bf 50}, 51 (1850).

\bibitem{98ZHCH}
L. Zhang, Y. Huang, J. Y. Chen and K. C. Hwang, Int. J.Fract. {\bf 92}, 325 (1998).



\end{thebibliography}
\end{document}